\documentclass[12pt,preprint]{aastex}

\shorttitle{Sensitivity of Stellar Evolution Models}
\shortauthors{Dotter et al.}

\newcommand{\Ms}{\mathrm{M_{\odot}}}
\newcommand{\Ls}{\mathrm{L_{\odot}}}
\newcommand{\Rs}{\mathrm{R_{\odot}}}
\newcommand{\Teff}{\mathrm{T_{eff}}}
\newcommand{\feh}{\mathrm{[Fe/H]}}

\begin{document}

\title{Stellar Population Models and Individual Element Abundances I: 
Sensitivity of Stellar Evolution Models}

\author{Aaron Dotter and Brian Chaboyer}
\affil{Department of Physics and Astronomy, Dartmouth College, 6127 Wilder Laboratory, Hanover, NH 03755}
\author{Jason W. Ferguson}
\affil{Physics Department, Wichita State University, Wichita, KS 67260-0032}
\author{Hyun-chul Lee and Guy Worthey}
\affil{Department of Physics and Astronomy, Washington State University, Pullman, WA 99164-2814}
\author{Darko Jevremovi\'c\altaffilmark{1} and E. Baron\altaffilmark{2}}
\affil{Homer L. Dodge Department of Physics and Astronomy, University of Oklahoma, 440 West Brooks, Room 100, Norman, OK 73019-2061}%\email{baron@nhn.ou.edu, djc@nhn.ou.edu}}
\altaffiltext{1}{Current Address: Astronomical Observatory, Volgina 7, 11160 Belgrade, Serbia}
\altaffiltext{2}{Computational Research Division, Lawrence Berkeley National Laboratory, MS 50F-1650, 1 Cyclotron Rd, Berkeley, CA 94720}

\begin{abstract}
Integrated light from distant galaxies is often compared to stellar population models via the equivalent widths of spectral features---spectral indices---whose strengths rely on the abundances of one or more elements.  Such comparisons hinge not only on the overall metal abundance but also on relative abundances.  Studies have examined the influence of individual elements on synthetic spectra but little has been done to address similar issues in the stellar evolution models that underlie most stellar population models.  Stellar evolution models will primarily be influenced by changes in opacities. In order to explore this issue in detail, twelve sets of stellar evolution tracks and isochrones have been created at constant heavy element mass fraction Z that self-consistently account for varying heavy element mixtures.  These sets include scaled-solar, $\alpha$-enhanced, and individual cases where the elements C, N, O, Ne, Mg, Si, S, Ca, Ti, and Fe have been enhanced above their scaled-solar values.  The variations that arise between scaled-solar and the other cases are examined with respect to the H-R diagram and main sequence lifetimes.
\end{abstract}


\section{Introduction}
In the first of a series of papers, the influence of individual elements on stellar parameters as they apply to H-R diagram morphologies and main sequence lifetimes are explored.  The next paper in this series, Lee et al. (2007, in preparation), will extend the analysis to spectral features and integrated light models based on the findings presented here.

Such an analysis is important because little is currently known about how an arbitrary heavy element mixture will influence stellar evolution in terms of, for example, the effective temperature of the main sequence turnoff or the red giant branch.  In particular, the consequences of altering the abundance of one heavy element with respect to others is largely untouched in the literature. Early studies focused on oxygen since it comprises about half of all metals by either number or mass fraction.  For example, \citet{vand} computed stellar evolution tracks with enhanced levels of oxygen but did not account for enhanced oxygen in the opacities.  He found that at low metallicity oxygen's influence on the nuclear reactions was more pronounced than on opacity, though again no oxygen rich opacities were available at the time. More recently, \citet{vandb} studied the influence of enhanced oxygen, both separately and as a part of $\alpha$-enhancement, in very metal poor stars ($\feh$=-2.27). These authors found results consistent with previous work that enhancing oxygen by 0.3 dex reduces age by 1 Gyr when the main sequence turn off luminosity is held constant.

The appearance of low temperature opacities with molecular absorption by \citet{alex} and the subsequent release of both scaled-solar and $\alpha$-enhanced mixtures marked an important improvement for stellar structure and evolution calculations.  Likewise with the introduction of improved high temperature opacities from the Opacity Project \citep{seat} and OPAL \citep{opal}. It has since become commonplace for isochrone libraries to include $\alpha$-enhanced isochrones computed self-consistently with both high and low temperature opacities \citep[for example]{kim,piet,salasnich,vand2}.

\citet{degl} used OPAL opacities computed for sixteen different heavy element mixtures, including different solar abundance determinations; cases where Z was comprised entirely of C, O, or Ne; and cases where certain elements and groups of elements were enhanced with respect to a standard solar mixture.  \citet{degl} focused on measuring the effects of the opacity variations on Li depletion during the pre-main sequence phase for stars with M=0.8, 1.0 and 1.2 $\Ms$.  These authors found that increasing the proportions of elements heavier than O at constant total Z does indeed increase Li depletion on the pre-main sequence.  However, their models did not include corresponding variations in either low temperature opacities or equation of state.  While it may be the case that specific metal abundances do not significantly influence the equation of state, the same cannot be said for low temperature opacities as will be demonstrated below.

\citet{weiss} examined the effects of maximizing abundance ratios in an $\alpha$-enhanced heavy element mixture at super-solar metallicity where differences in the makeup of Z are most pronounced.  In addition to comparing recent low-temperature opacity calculations to the previous generation, \citet{weiss} found that ``the lifetimes of $\alpha$-enriched low mass stars are sensitive to the individual abundance pattern" and thus ``accurate age determinations ... may require very detailed knowledge of the chemical composition''

It is well known \citep{trip,korn,serv} that increasing the abundance of a particular element can alter a synthetic spectrum by increasing or decreasing the absorption features, or opacity, of various atomic or molecular lines.  Changes in opacity will in turn alter stellar evolution calculations.  To constrain the analysis to relative abundances, all stellar evolution calculations were performed at constant heavy element mass fraction Z consistent with a calibrated solar model.  It is also true that nuclear reaction rates for the CNO cycle are sensitive to the absolute and relative abundances of C, N, and O.  However, the results presented below indicate that any changes rely almost entirely on changes to the opacity.

The purpose of this study is to explore the influence that several of the most abundant metals have on the properties of stellar evolution models relative to scaled-solar abundances.  The paper proceeds as follows, $\S$2 describes the details of how the elements were chosen and the heavy element mixtures created, $\S$3 outlines the opacity and stellar evolution computations, $\S$4 presents the mean opacities, isochrones, and stellar lifetimes, $\S$5 provides some analysis of these findings, and $\S$6 summarizes this work and identifies possibilities for further exploration.

\section{Abundances}
Ten elements were selected for study based on their overall importance to the metal content of the Sun and other stars.  Oxygen and carbon are the biggest contributors to the total metal content in stars and, along with nitrogen, contribute to nuclear reactions in stars through the CNO cycle.  Oxygen is also the first $\alpha$-capture element. The bulk of the remaining selections are additional $\alpha$-capture elements: Ne, Mg, Si, S, Ca, and Ti.  Argon has been left unchanged from its solar value in this study because it is a noble gas and has an abundance well below neon.  The final element chosen for study is iron because of its importance as an endpoint to nucleosynthesis, its relatively high abundance in the Sun and other stars, and its importance to opacity \citep{opal}.  In addition to the individually enhanced elements an $\alpha$-enhanced mixture, where the abundances of O, Ne, Mg, Si, S, Ca, and Ti were each enhanced by the same amount, has been included in the analysis.

The purpose of this paper is to identify what, if any, significant differences in stellar parameters arise by enhancing the abundance of one element at constant metal mass fraction. To achieve this goal all stellar evolution calculations were performed at constant mass fractions of hydrogen (X), helium (Y), and overall metals (Z) except in special cases where the Z value has been noted.  In general, the only difference between the models is the makeup of Z.  The choice to hold X, Y, and Z constant was made to reduce the variations incurred by changing either X or Y in response to changing Z and also to minimize the amount and scope of additional input physics required for each mixture (see section $\S$3 for details). It is important to note that the act of enhancing one metal at constant Z must be done at the expense of all the other metals.  For the more abundant metals, primarily oxygen, the deviations from scaled-solar can be strongly influenced by the depletion of other elements (see $\S$4.1.2 and Figure \ref{opac3}). Because C, N, and O are three of the most abundant metals and play a direct role in the evolution of low mass stars, additional sets of models have been calculated for the same C-, N-, and O-enhanced mixtures but with $\feh$=0 (\emph{not constant Z}). The mass fractions used in all calculations are discussed in $\S$3.

%Figure 1
\clearpage
\begin{figure}
\epsscale{0.8}
\plotone{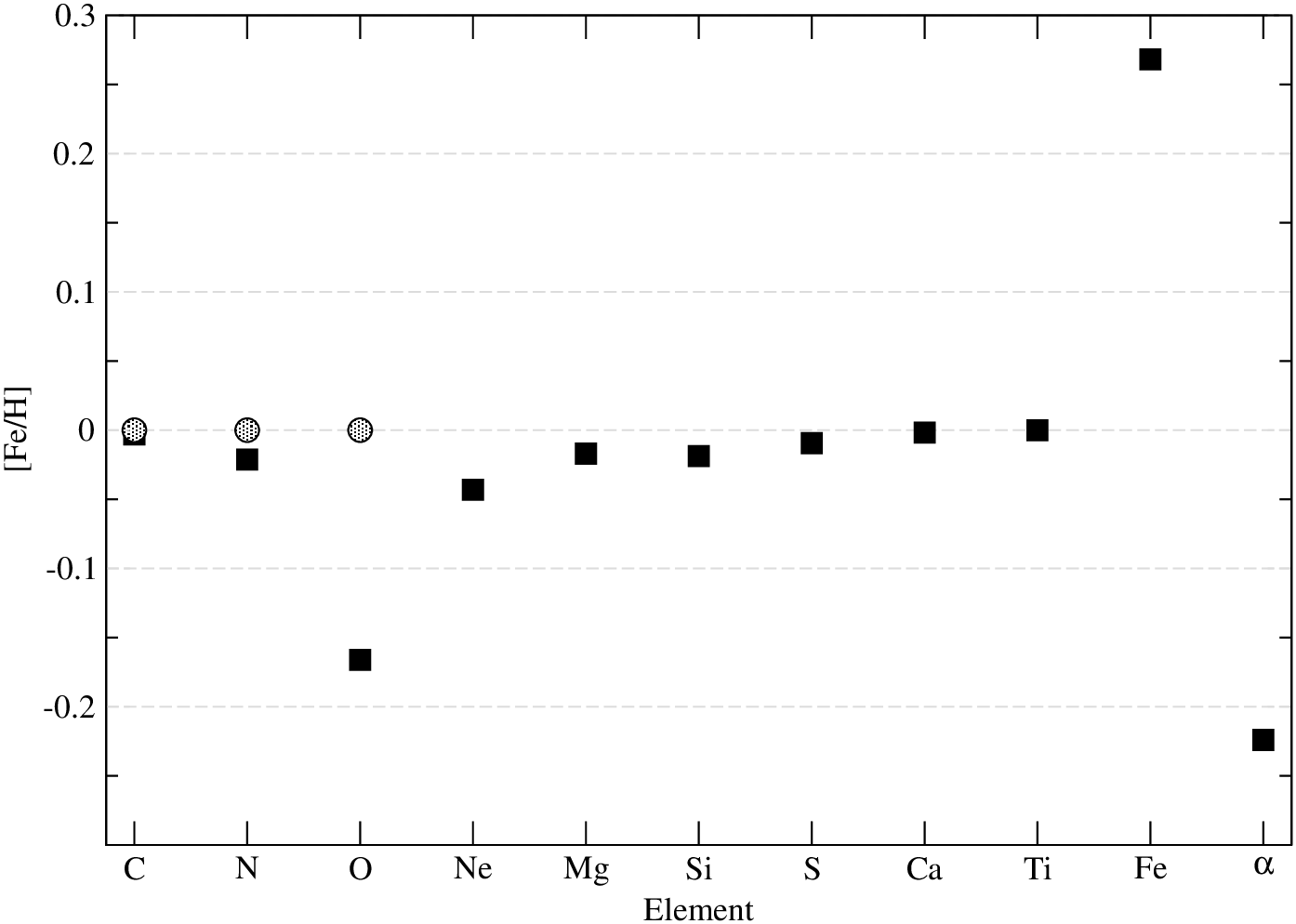}
\caption{The figure demonstrates the effect of varying the heavy element mixture at constant Z.  The scaled-solar case is assumed to have $\feh$=0.  The filled squares represent the constant Z mixtures while the shaded circles represent C-, N-, and O-enhancement at $\feh$=0.\label{feh}}
\end{figure}
\clearpage

Figure \ref{feh} displays the $\feh$ values for each of the mixtures considered. Filled squares represent the constant Z cases while the shaded circles represent the C-, N-, and O-enhanced at $\feh$=0 cases.  At constant Z, only the O-, Fe-, and $\alpha$-enhanced mixtures depart by more than 0.05 dex from the scaled-solar mixture which is assumed to have $\feh$=0.  With the exception of the Fe-enhanced case, all unenhanced elements follow the trend shown in Figure \ref{feh} for [Fe/H].

The adopted scaled-solar mixture is that of \citet[hereafter GS98]{gs98}.  Beginning with the GS98 abundance list, the eleven additional heavy element mixtures were created by separately enhancing C by 0.2 dex and each of the remaining elements by 0.3 dex above its scaled-solar value. The $\alpha$-enhanced mixture was created by simultaneously enhancing O, Ne, Mg, Si, S, Ca, and Ti by 0.3 dex above their scaled-solar values; it was included because such mixtures are commonplace in many isochrone libraries and because it allows for a direct comparison of $\alpha$-enhancement with each of its components.    

In the remainder of this paper, each mixture will be referred to by either scaled-solar, $\alpha$-enhanced, or the chemical symbol of the enhanced element.

\section{Opacities and Stellar Evolution Models}

Stellar evolution computations were carried out with the Dartmouth Stellar Evolution Program (DSEP).  DSEP has been modified to use the detailed equation of state code FreeEOS\footnote{http://freeeos.sourceforge.net/} \citep{irwin} that explicitly accounts for the heavy element mixture.  Many other details of DSEP can be found in \citet{chab,bjor}.  Convective core overshoot is treated using the method developed by \citet{dema}.  Specifically, overshoot is linearly ramped from 0.05 pressure scale heights at the minimum stellar mass where the convective core appears to 0.2 pressure scale heights at 0.2 $\Ms$ or more above the minimum.  Most of the models employ the Eddingtion T-$\tau$ surface boundary condition.  However, stellar evolution calculations with PHOENIX model atmosphere \citep{phxa,phxb} boundary conditions were performed for the scaled-solar and $\alpha$-enhanced cases as a check on the Eddington T-$\tau$ boundary condition results.  See Figure \ref{iso6} and $\S$4.3 for the boundary condition comparison.

Both high and low temperature Rosseland mean opacity tables were constructed for the heavy element mixtures.  High temperature opacities by \citet{opal} were obtained from the OPAL website\footnote{http://www-phys.llnl.gov/Research/OPAL/}.  Due to their computationally intensive nature, low temperature opacities \citep{ferg} were computed for small windows surrounding the mass fractions of interest.  Specifically, low-temperature opacities were calculated for hydrogen mass fractions X=0.7 and 0.8 with Z=0.01 and 0.02 for each X. Since the stellar evolution calculations include the effects of diffusion and gravitational settling it was important to include a range of both X and Z rather than adopting one value for each. The transition between the \citet{ferg} low temperature opacities and OPAL high temperature opacities is carried out between Log T=4.0 and 4.1.  Agreement between the two sets of tables in the overlap region is discussed at length by \citet{ferg}.

All stellar evolution computations were performed with a solar-calibrated mixing length $\alpha_{MLT}$=1.83 and initial composition X$_i$=0.707 and Z$_i$=0.0188 except for the C-, N-, and O-enhanced models computed at $\feh$=0 that have Z=0.0221, 0.0198, and 0.0276, respectively.  The initial composition and mixing length adopted are required to obtain a calibrated solar model with DSEP using the configuration described above.  Solar parameters used for calibration purposes were the same as adopted by \citet{bahc}: $\Ls$ = 3.8418 x 10$^{33}$ erg/s, $\Rs$ = 6.9598 x 10$^{10}$ cm, and surface (Z/X)$_{\odot}$ = 0.0229 (GS98) at $\tau_{\odot}$ = 4.57 Gyr.  A solar-calibrated mixing length meets each of the solar parameters just listed to 1 part in 10$^4$ or better at $\tau_{\odot}$.

Stellar evolution models with masses between 0.15 and 4 $\Ms$ were computed for each heavy element mixture, allowing for analysis of ages as young as a few hundred Myr.  Evolution was followed from the pre-main sequence to the tip of the red giant branch for M $<$ 2 $\Ms$ and up to the onset of thermal pulses on the asymptotic giant branch for the upper mass range.  The analysis is limited to the first ascent of the giant branch in this paper.

DSEP explicitly tracks elements up to and including oxygen.  The abundances of the enhanced elements from Ne to Fe are explicitly included in the opacity tables and the EOS but are assumed to maintain the same relative values throughout the evolution.

\section{Results}
The most important variable when considering heavy element abundance variations in the stellar evolution calculations is the opacity.  Therefore the first part of this section is devoted to a comparison of opacities and a discussion of the reasons for any major differences between enhanced-element and scaled-solar opacities.

\subsection{Opacities}

\subsubsection{Low temperature opacities}
\clearpage
%Figure 2
\begin{figure}
\epsscale{0.8}
\plotone{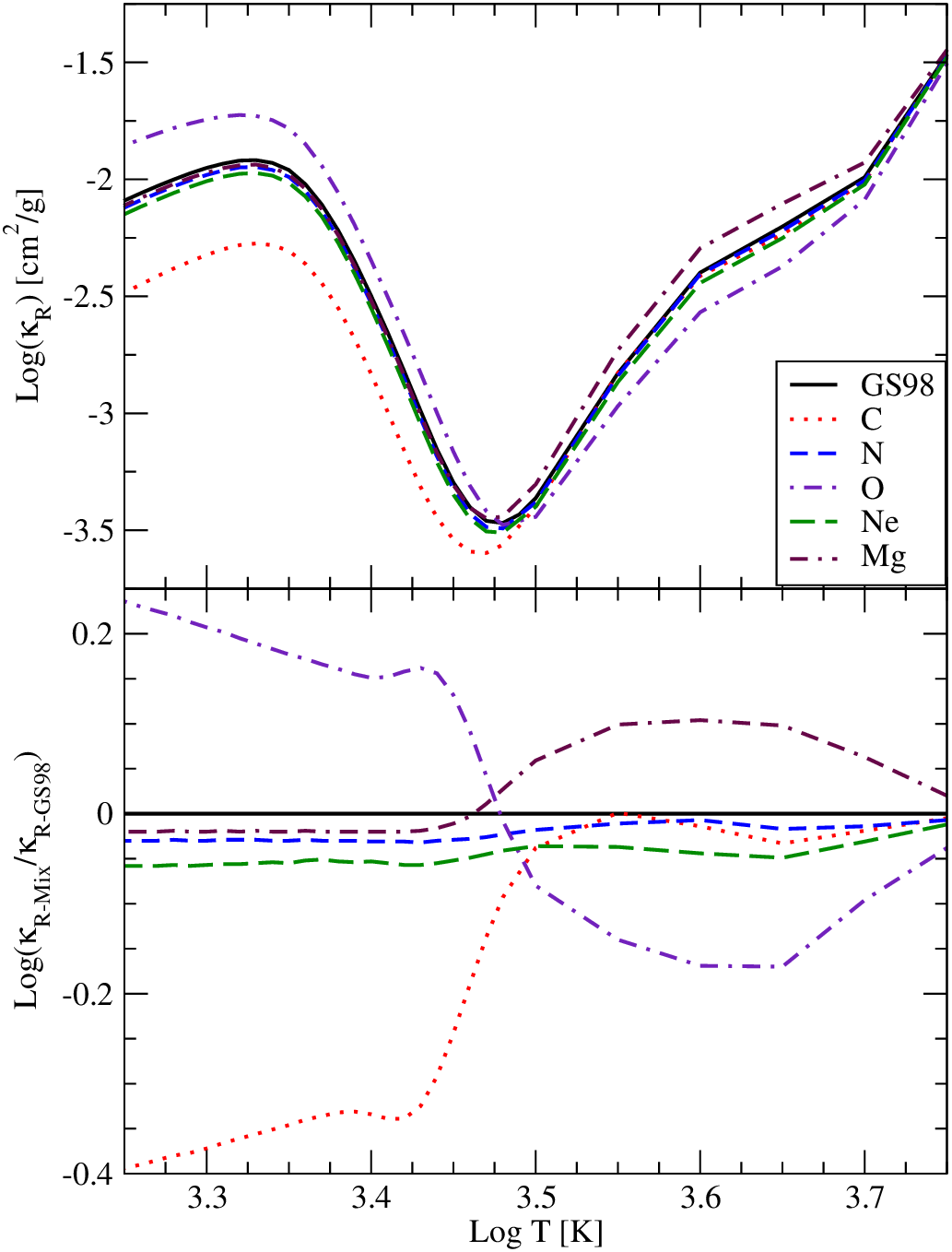}
\caption{Comparison of enhanced C, N, O, Ne, and Mg opacities to scaled-solar in both absolute value (upper panel) and in ratio to scaled-solar (lower panel), all at Log R=-1.5. 'GS98' refers to scaled-solar while 'Mix' refers to the individual elements.  The stellar evolution models exist above Log T=3.5, given this constraint Mg and O exhibit the most dramatic changes from scaled-solar while the remaining elements differ only slightly.\label{opac1}}
\end{figure}
%Figure 3
\begin{figure}
\epsscale{0.8}
\plotone{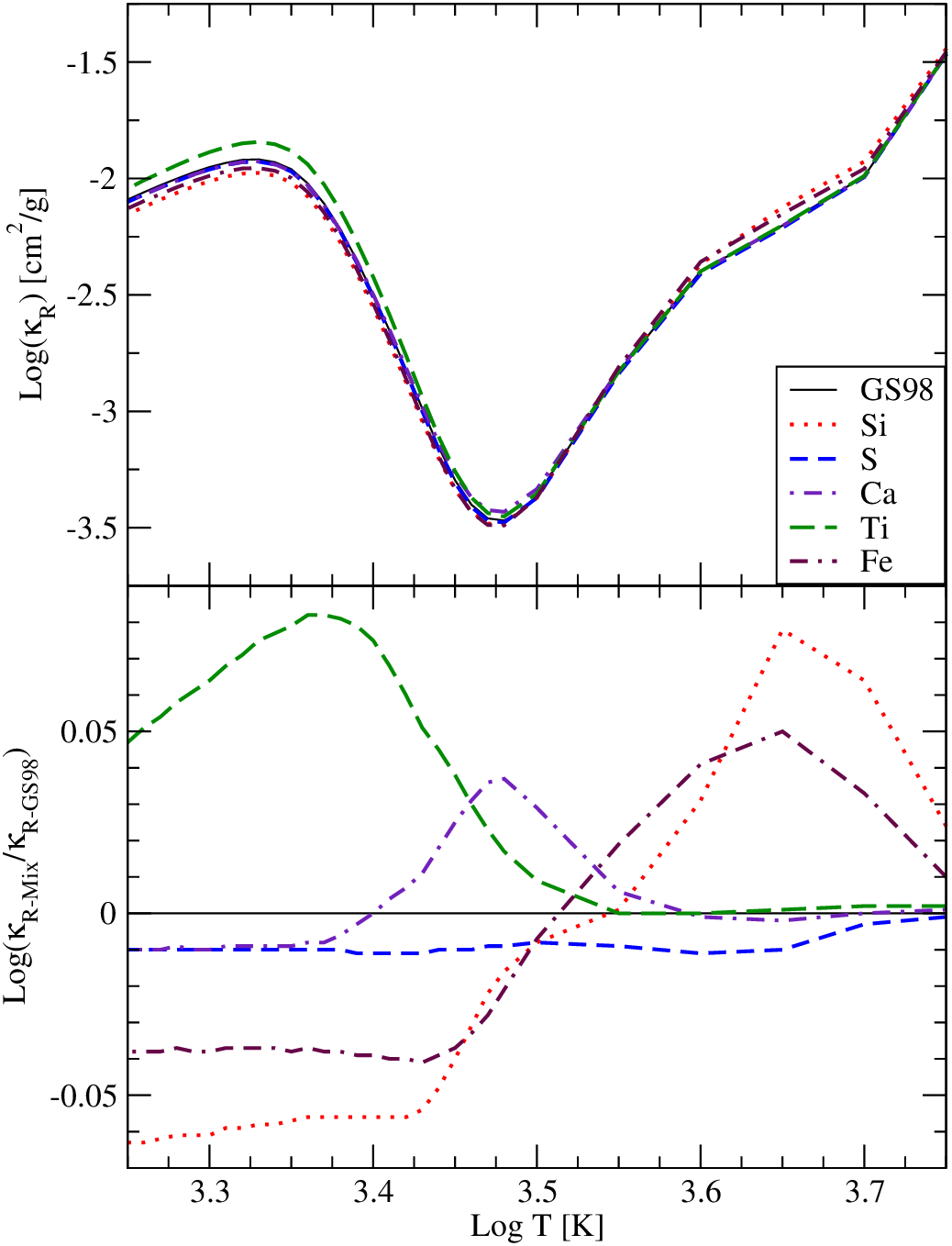}
\caption{Same as Figure \ref{opac1} for Si, S, Ca, Ti, and Fe but note the change in scale for the lower panel. For Log T=3.5 and above, Si and Fe exhibit the largest increases over scaled-solar (GS98) though both are less substantial than Mg in Figure \ref{opac1}.\label{opac2}}
\end{figure}
\clearpage
Low temperature opacities were calculated using the machinery described by \citet{ferg}; the following discussion describes some of the most important changes that arose due to the new heavy element mixtures. 

Enhancing the abundance of a single element (while conserving the total amount of metals, Z) leads to some interesting and surprising changes in the Rosseland mean opacity at low temperatures (all further references to mean opacity imply Rosseland mean opacity).  Figures \ref{opac1} and \ref{opac2} show comparisons between the mean opacity computed with solar abundances from GS98 and with the mean opacity from the individually enhanced abundance sets.  The figures show the mean opacity from Log T=3.25 to 3.75 at Log R=-1.5 (where R=$\rho$/T$_6^3$ and T$_6$ is the temperature in millions of K).  Between Log T=3.75 and 4 the opacity is dominated by H and He and changes very little with variations in the heavy element mixture.  

The largest changes occur for the most abundant elements.  For example, in Figure \ref{opac1} increasing C and O lower and raise the mean opacity, respectively, at temperatures where water is important (Log T $<$ 3.5).  As expected, an increase in C reduces the amount of O available for water and the mean opacity decreases.  Increase O and the amount of water opacity increases.  However, raising the abundance of Ne lowers the mean opacity.  The reduction occurs due to the conservation of the other metals, more Ne means less C, N, and O.  Surprisingly, raising Mg by a factor of two does make a significant change in the opacity for 3.45 $<$ log T $<$ 3.75.  The rise is due almost entirely to the strong Mg II line at 2798 $\AA$ that has very large wings.  This line is located near the peak of the weight function that defines the Rosseland mean for the temperature range in question and contributes significantly to the mean opacity.  In addition, Mg has strong continuum absorption in the IR at these temperatures, where continuum plays an important role in the mean opacity.

For heavier $\alpha$-enhanced elements (Figure \ref{opac2}) the story is much the same.  Si-enhancement raises the opacity through strong absorption by Si II for 3.55 $<$ log T $<$ 3.75 and lowers the mean opacity at lower temperatures as it competes with Ti for O.  S-enhancement causes a mild and fairly constant reduction in the low temperature opacity by displacing the other elements to maintain constant Z. Ca raises the opacity between 3.45 $<$ Log T $<$ 3.5 because atomic line transitions become important at those temperatures.  Likewise as Ti is increased its largest effect is at low temperatures where TiO opacity is lying on top of the water bump at Log T=3.35.  This effect is below the stellar effective temperatures under consideration here.  Enhancing Fe changes the mean opacity in similar ways as the Si comparison shows.  At moderate temperatures more Fe means more Fe II lines and an increase in the opacity, while at lower temperature more Fe means less O for water.  

\subsubsection{High temperature opacities}
\clearpage
%Figure 4
\begin{figure}
\epsscale{0.8}
\plotone{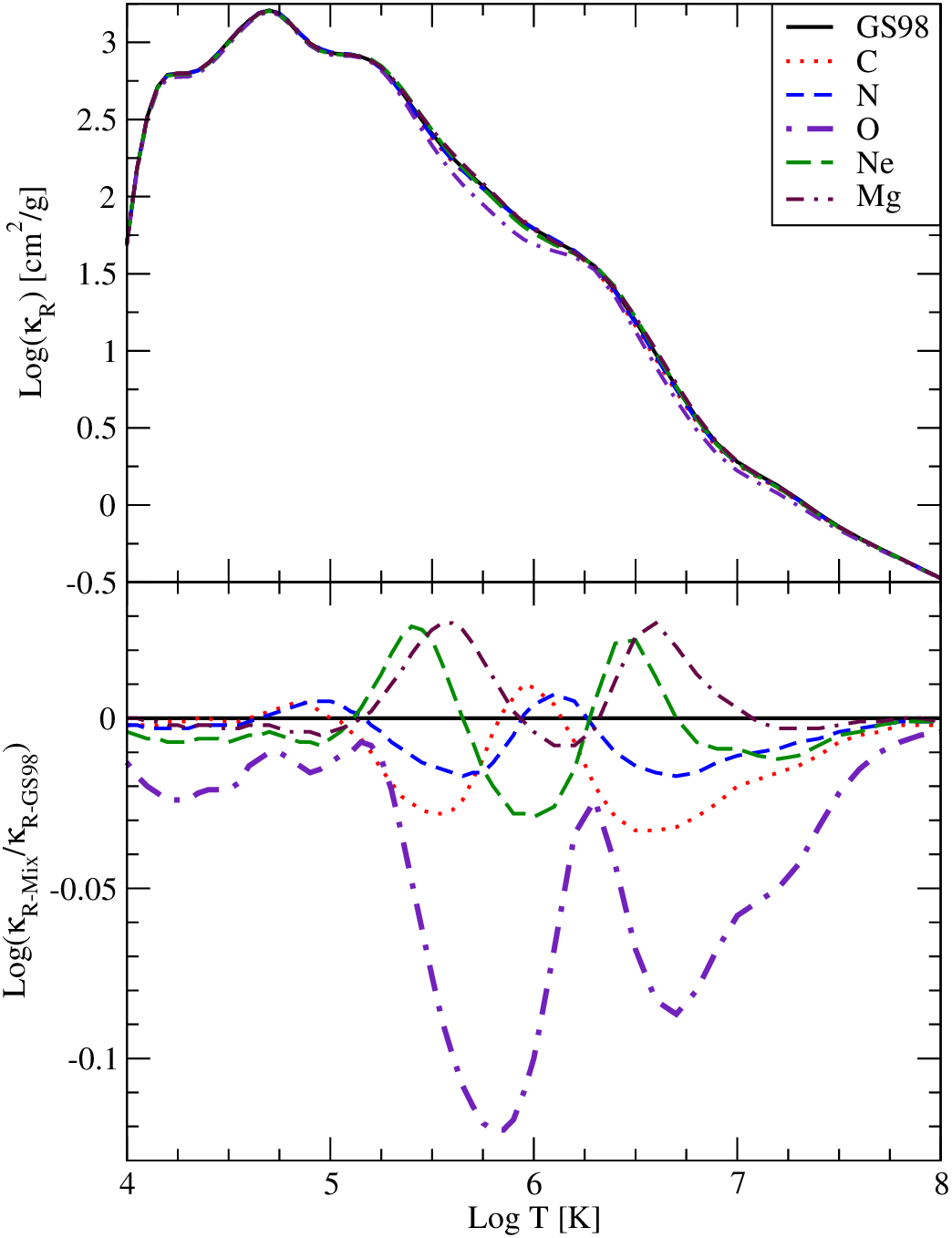}
\caption{Similar to Figure \ref{opac1}, this shows differences between scaled-solar (GS98) and the enhanced elements C, N, O, Ne, and Mg for OPAL opacities in absolute value (upper panel) and in ratio to scaled-solar (lower panel), all at Log R=-1.5. O shows by far the largest departure from scaled-solar due to it displacing the heavier elements.\label{opac3}}
\end{figure}
%Figure 5
\begin{figure}
\epsscale{0.8}
\plotone{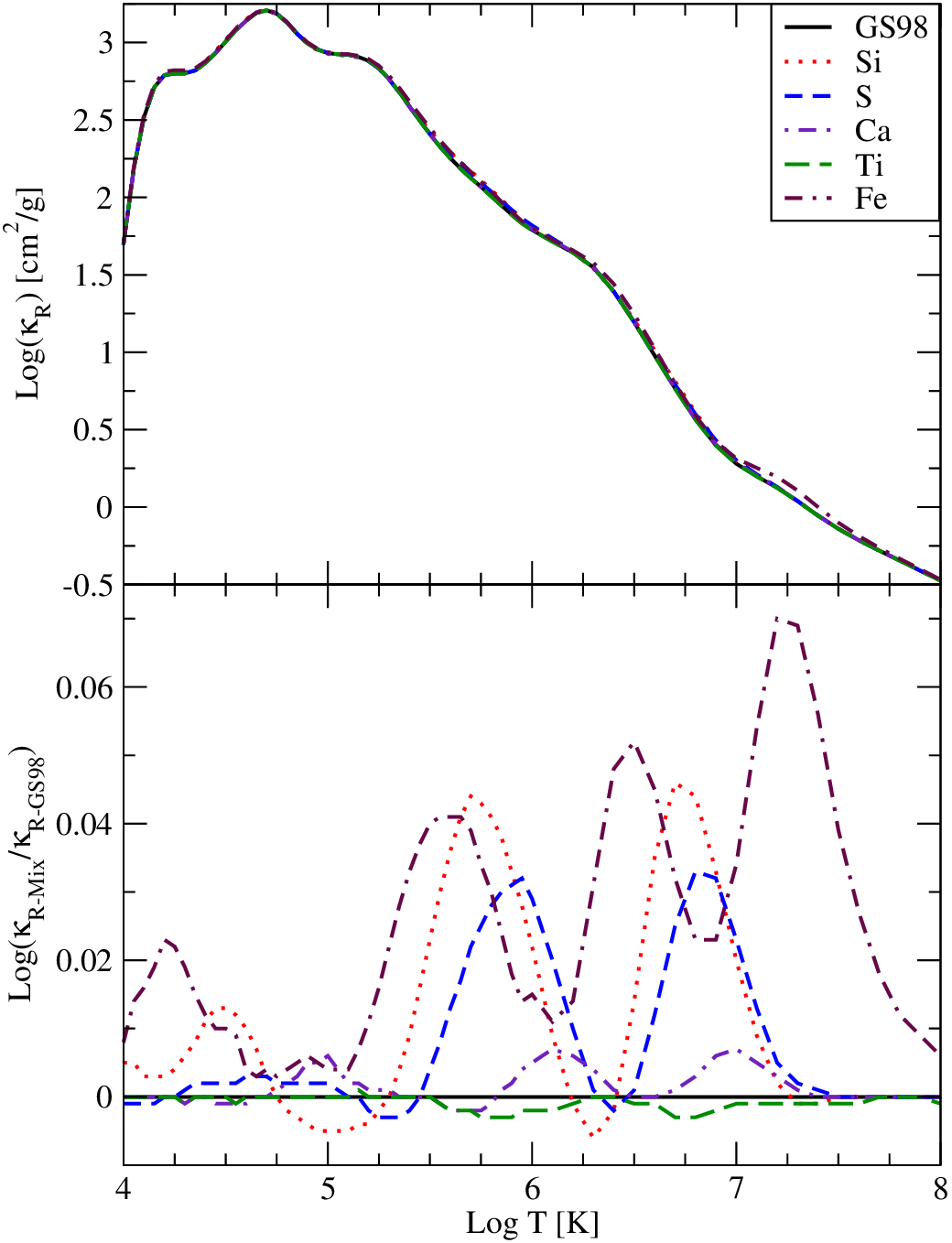}
\caption{Same as Figure \ref{opac3} for Si, S, Ca, Ti, and Fe. Note the difference in the lower panel between this and Figure \ref{opac3}. Enhanced Si, S, and Fe all show large bumps in opacity due to increased absorption while Ca and Ti contribute little due to their low abundances.\label{opac4}}
\end{figure}
\clearpage
 Figures \ref{opac3} and \ref{opac4} are the OPAL opacity counterparts of Figures \ref{opac1} and \ref{opac2}.  The figures show the full temperature range utilized by the stellar evolution code up to Log T=8 where the opacities converge at Log R=-1.5. The largest change to the mean opacity at high temperature occurs for O-enhancement due to its replacing large amounts of Ne, Mg, Si, S, and Fe.  Because many of the heavier elements (Ne, Mg, Si, S, Fe) increase the opacity between Log T=5 and 8, the consequence of enhancing C, N, and O at constant Z is generally to reduce the opacity relative to scaled-solar in this temperature range as the figures clearly demonstrate.  The elements Ne, Mg, Si, S, Fe, and to a lesser extent Ca, exhibit a series of bumps in the opacity relative to scaled-solar between Log T=5 and 6, 6 and 7, and (in the case of Fe) 7 and 8. The bumps are caused by increased absorption by the enhanced element in question near where the Rosseland mean weight function peaks for those temperatures (C. Iglesias, private communication 2006). The differences in the locations and strengths of the bumps are due to different excitation levels and ionization potentials.

\subsection{Isochrones}
In order to span a range of ages and stellar masses, isochrones were computed for six ages: 0.5, 1, 2, 4, 8, and 12 Gyr.  The isochrones are presented in the theoretical H-R diagram because this investigation is concerned with theoretical parameters such as age, luminosity, and $\Teff$. Colors, magnitudes, and other spectral properties will be addressed in a future paper.
\clearpage
%Figure 6
\begin{figure}
\epsscale{1.0}
\plotone{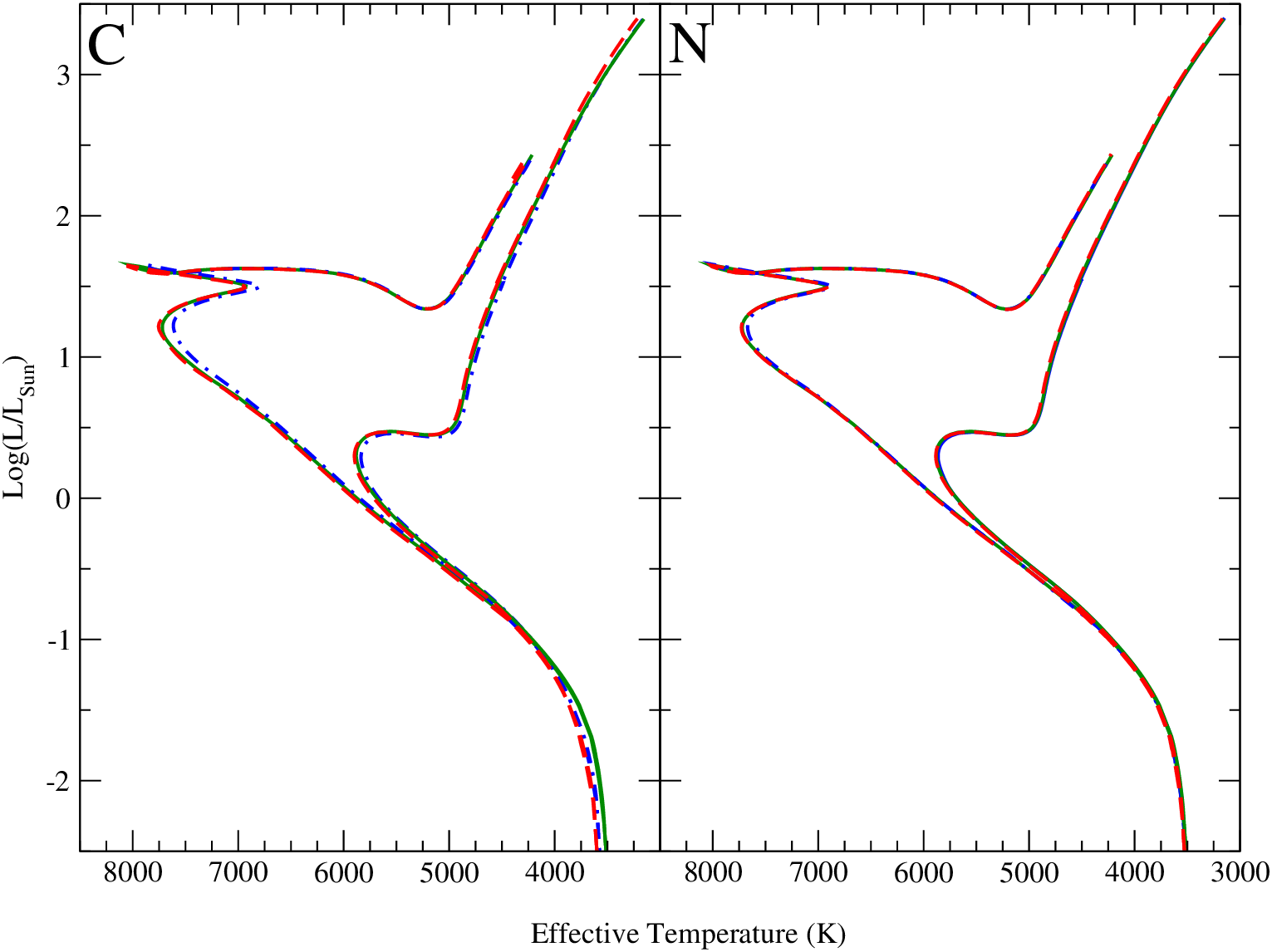}
\caption{Comparisons between scaled-solar (solid lines) and the enhanced elements C and N at constant Z (dashed lines) and $\feh$=0 (dot-dashed lines) for ages 1 and 8 Gyr.}
\label{iso1}  
\end{figure}
%Figure 7
\begin{figure}
\epsscale{1.0}
\plotone{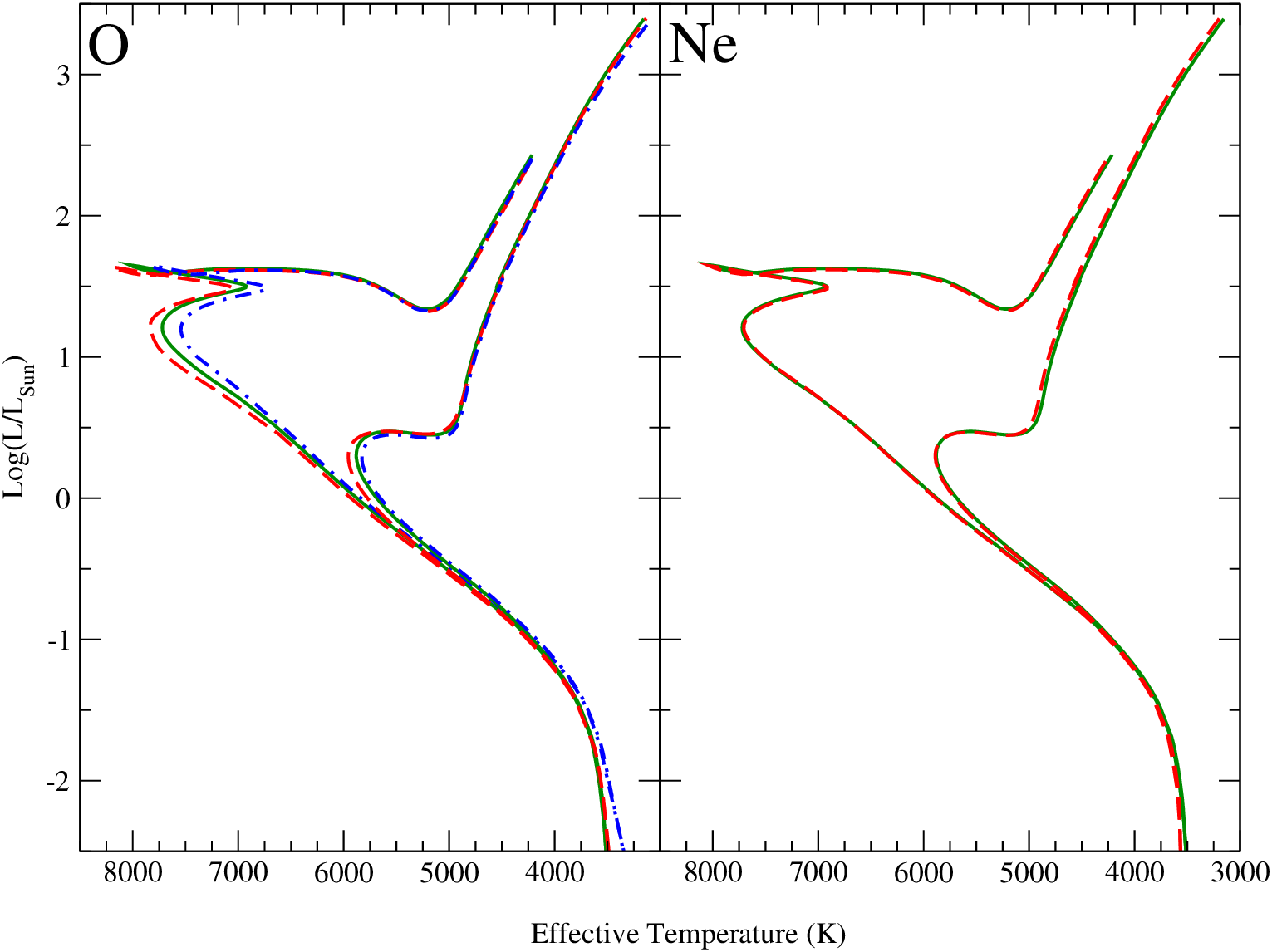}
\caption{Same as Figure \ref{iso1} for enhanced elements O and Ne. In the O-enhanced plot there is an additional set of isochrones for which [Fe/H]=0 (dot-dashed lines).}
\label{iso2}  
\end{figure}
%Figure 8
\begin{figure}
\epsscale{1.0}
\plotone{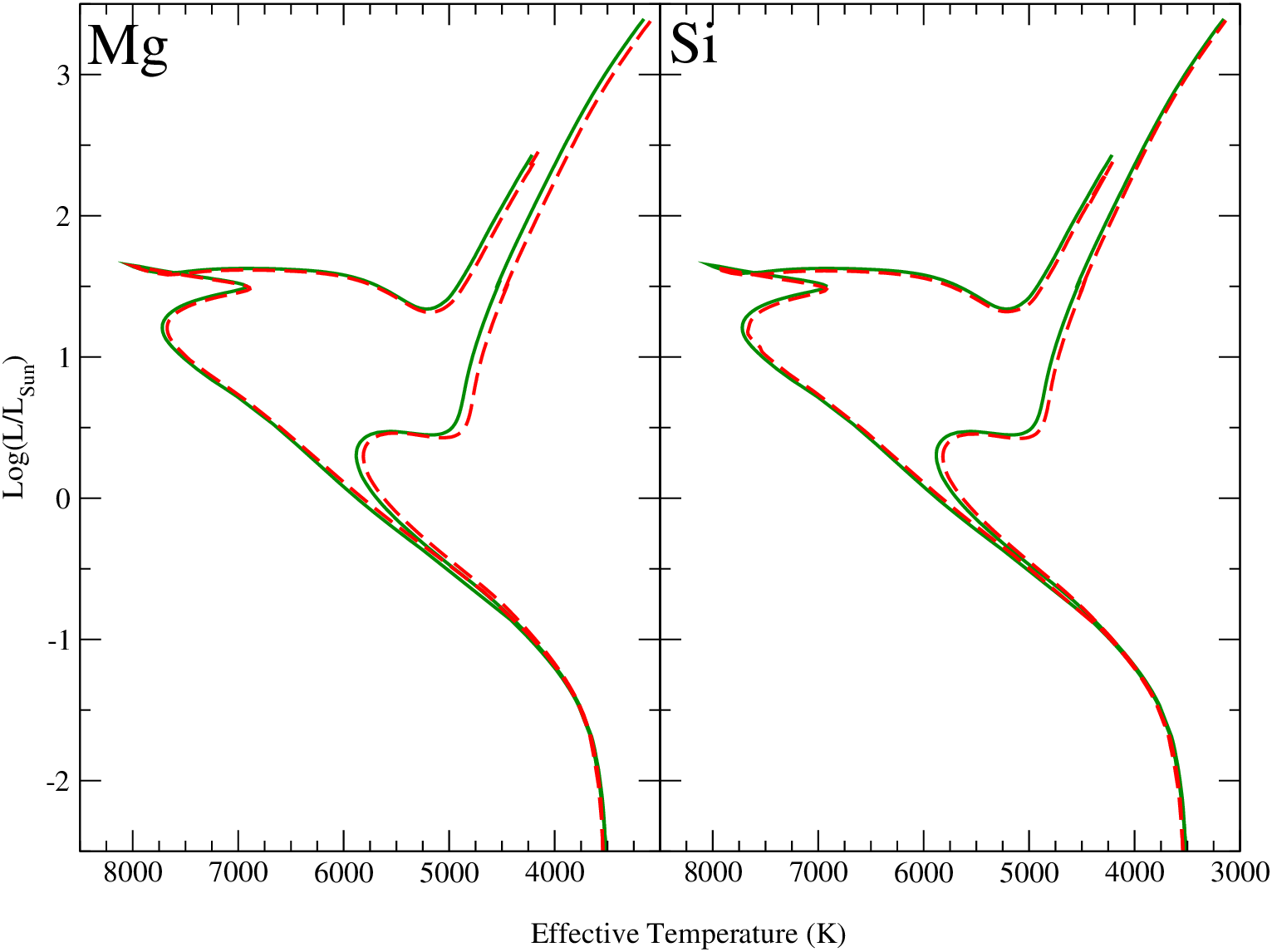}
\caption{Same as Figure \ref{iso1} for enhanced elements Mg and Si. Note the cooler temperatures on the main sequence and red giant branch at older ages.}
\label{iso3}  
\end{figure}
%Figure 9
\begin{figure}
\epsscale{1.0}
\plotone{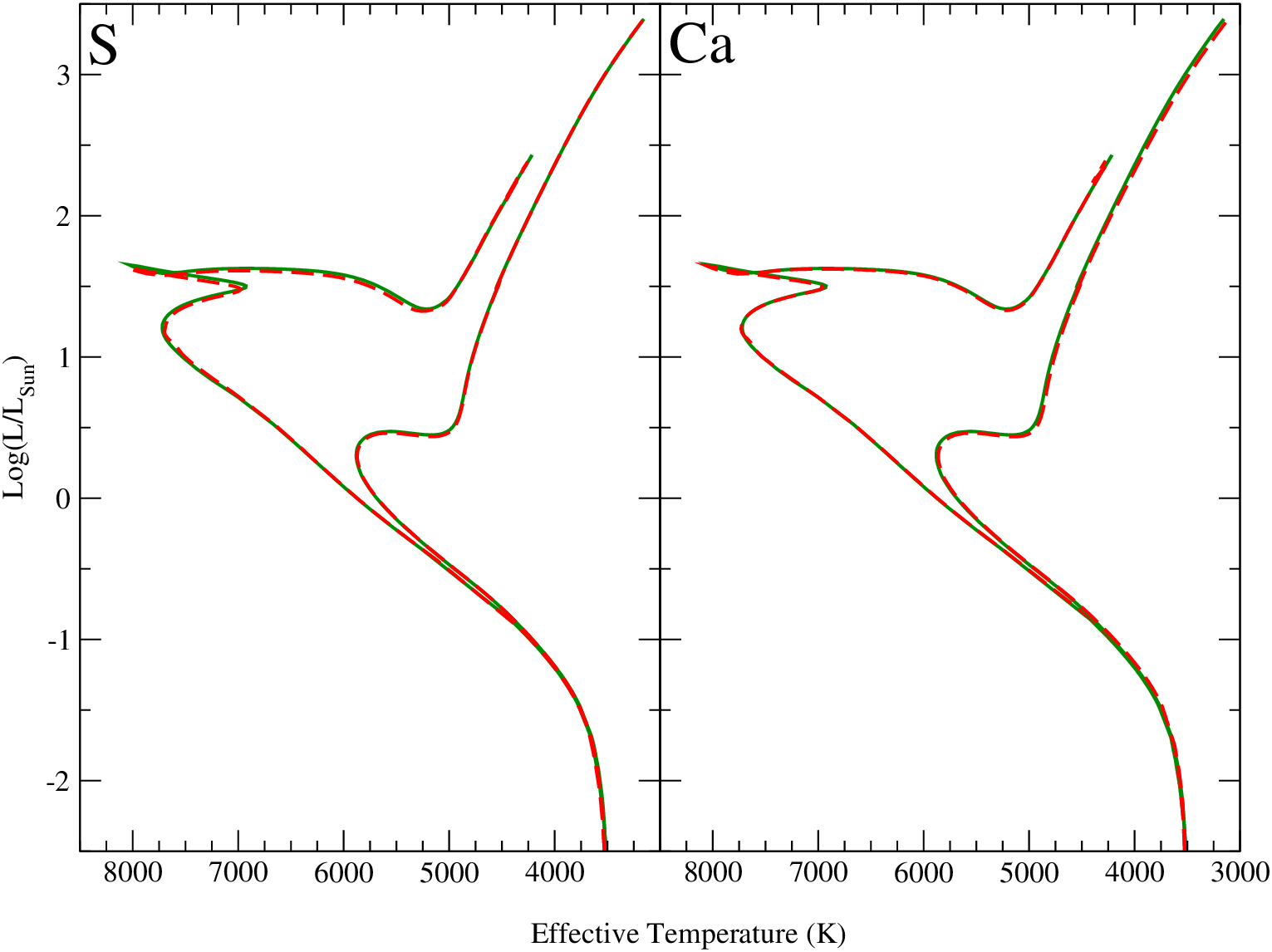}
\caption{Same as Figure \ref{iso1} for enhanced elements S and Ca.}
\label{iso4}  
\end{figure}
%Figure 10
\begin{figure}
\epsscale{1.0}
\plotone{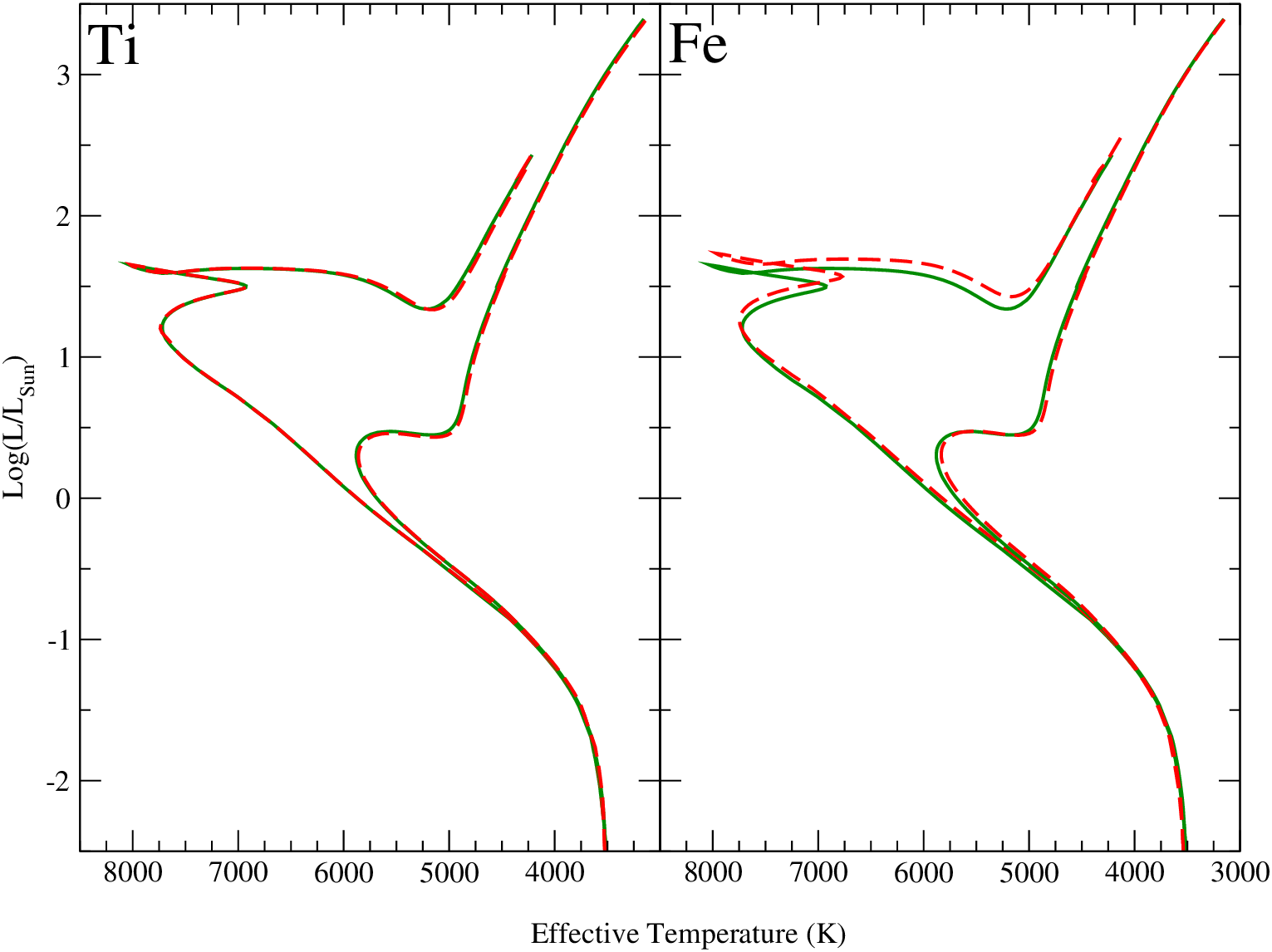}
\caption{Same as Figure \ref{iso1} for enhanced elements Ti and Fe. Note the increased luminosity of the sub-giant branch at 1 Gyr for the Fe-enhanced case.}
\label{iso5}  
\end{figure}

%Figure 11
\begin{figure}
\epsscale{1.0}
\plotone{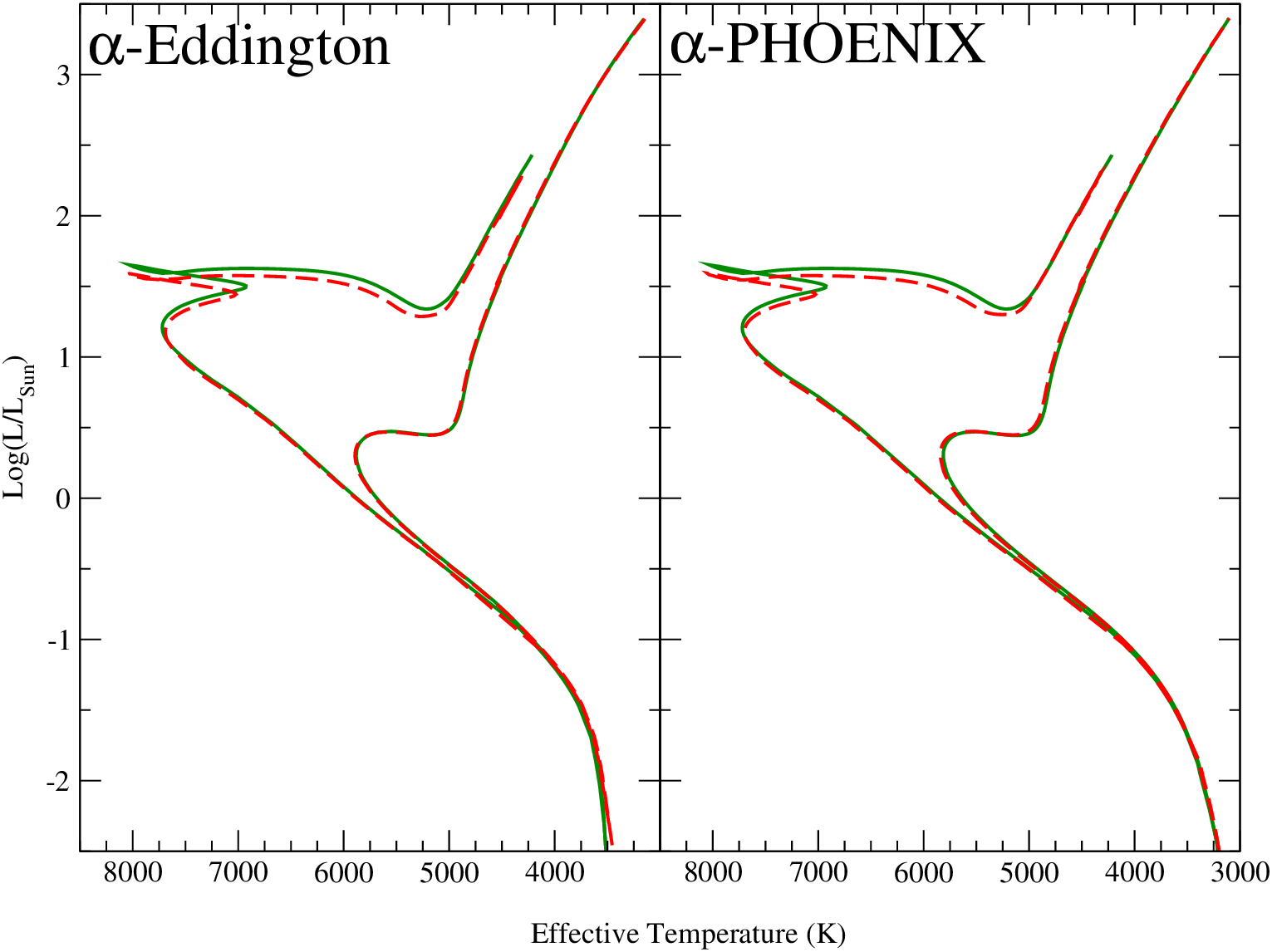}
\caption{Same as Figure \ref{iso1} for $\alpha$-enhancement with both Eddington and PHOENIX boundary conditions. Differentially, the results between boundary conditions are found to be quite similar. See $\S$4.3 for discussion.}
\label{iso6}  
\end{figure}
\clearpage
Figures \ref{iso1} through \ref{iso6} show 1 and 8 Gyr isochrones with scaled-solar always represented by the solid line and the enhanced element (noted in the upper left of each panel) by the dashed line. For C, N, and O (Figures \ref{iso1} and \ref{iso2}) the isochrones computed at $\feh$=0 are represented by dot-dashed lines. Figure \ref{iso1} compares scaled-solar to enhanced C and N, Figure \ref{iso2} to O and Ne, \ref{iso3} to Mg and Si, \ref{iso4} to S and Ca, \ref{iso5} to Ti and Fe, and \ref{iso6} to $\alpha$-enhanced isochrones made with the Eddingtion T-$\tau$ and PHOENIX boundary conditions. The boundary condition comparison will be discussed further in $\S$4.3. 

Differential isochrone comparisons are listed in a series of six tables (corresponding to the six ages listed above) containing about ten points from the lower main sequence to the tip of the red giant branch, including the main sequence turn off, for each age.  The tables are organized by first printing the absolute luminosity and $\Teff$ from the scaled-solar isochrones in the first two columns, followed by differences in $\Teff$ at constant luminosity between scaled-solar and the other cases in the remaining columns.  In most cases the luminosity is more or less constant with respect to change in composition.  The only exceptions are Fe- and $\alpha$-enhanced (further discussion can be found in $\S$5). At younger ages the turn off and sub-giant regions are slightly more luminous than in the scaled-solar case for Fe-enhanced while the opposite is true for $\alpha$-enhanced. Note that these cases show the strongest deviations in $\feh$ from scaled-solar in Figure \ref{feh}. Nevertheless, luminosity was chosen as the constant factor amongst all the isochrones. The differences in $\Teff$ are reported in the sense that $\Delta \Teff$(mix) = $\Teff$(mix) -- $\Teff$(scaled-solar).

\begin{figure}
\epsscale{1.0}
\plotone{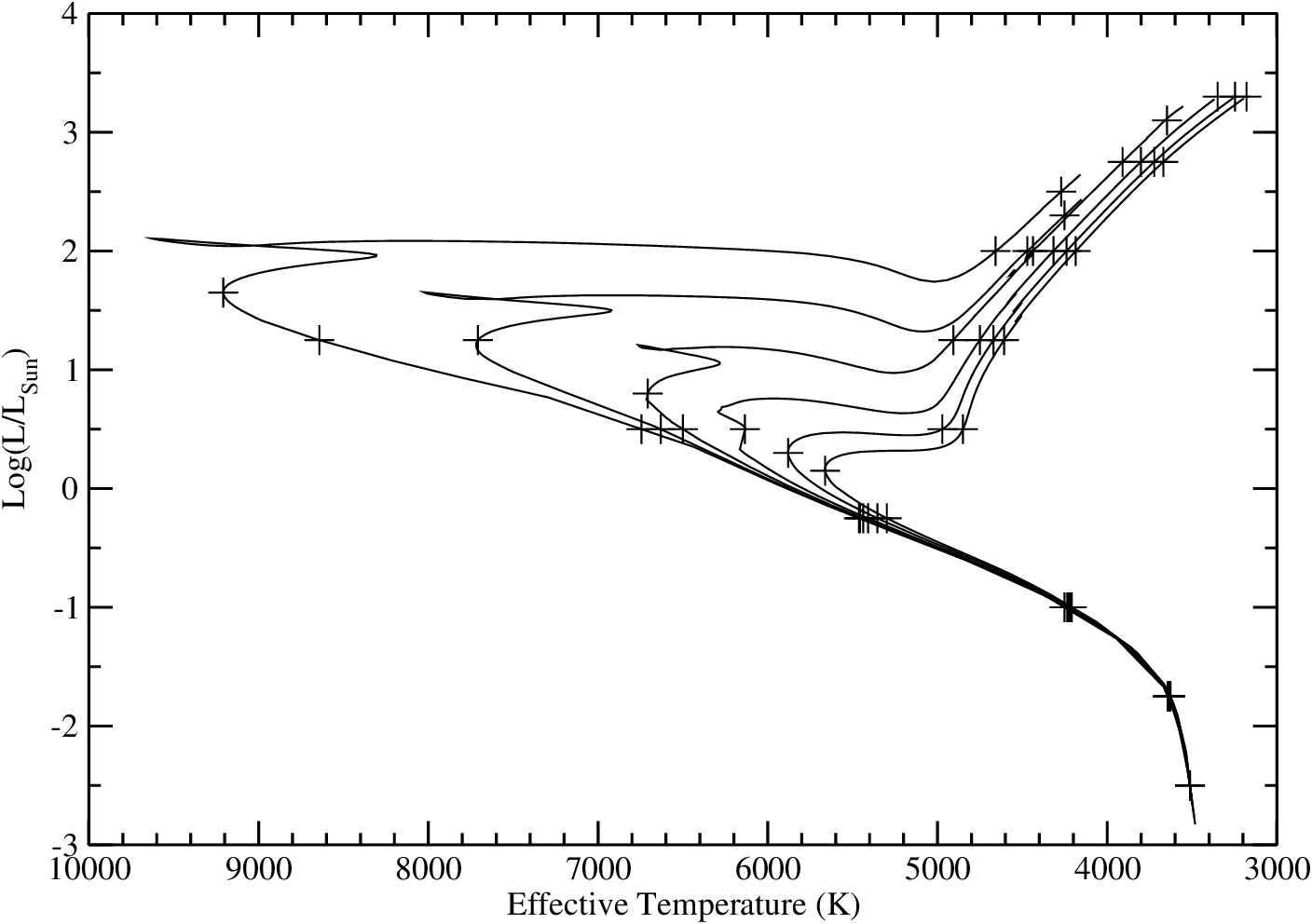}
\caption{The locations of the tabulated points (crosses) in the H-R diagram overlaid on the scaled-solar isochrones for ages 0.5, 1, 2, 4, 8, and 12 Gyr (left to right).  It is at these points that the isochrones for different heavy element mixtures are compared to scaled-solar.  See Tables \ref{tab1} through \ref{tab6} for the comparisons.}
\label{iso_table}
\end{figure}

For reference, Figure \ref{iso_table} shows the location of the tabulated points (see Tables \ref{tab1} through \ref{tab6}) on the scaled-solar isochrones.  The points have been chosen to avoid areas in the H-R diagram where $\Teff$ is changing rapidly, specifically the hook for the younger isochrones and the sub-giant branch for all ages.  These are the locations where, for each age, the effective temperatures are compared. Note that all tabulated points more luminous than the turn off point (the hottest point in each table) are assumed to be on the red giant branch.  At younger ages it is possible for some luminosities to occur more than once on an isochrone. In these cases the cooler point has always been used in the analysis.
\clearpage
%Figure 12
\clearpage
% %TABLE 1
\begin{deluxetable}{cccccccccccccccc}
\tablecolumns{16}
\tablewidth{0pc}
\tablecaption{Differences in the H-R Diagram at 500 Myr\label{tab1}}
\tablehead{
\multicolumn{2}{c}{Scaled-Solar}&\multicolumn{11}{c}{$\Delta \Teff$ (K) at Constant Log(L/$\Ls$)}\\
\colhead{Log(L/$\Ls$)}&\colhead{$\Teff$ (K)}&\colhead{C$^a$}&\colhead{C$^b$}&\colhead{N$^a$}&\colhead{N$^b$}&\colhead{O$^a$}&\colhead{O$^b$}&\colhead{Ne}&\colhead{Mg}&\colhead{Si}&\colhead{S}&\colhead{Ca}&\colhead{Ti}&\colhead{Fe}&\colhead{$\alpha$}}
\startdata
%L/Lo   Teff      Ca       Cb     Na      Nb      Oa      Ob     Ne      Mg      Si       S      Ca      Ti      Fe   alpha
 2.50 & 4331 &    11 &   -18 &     8 &    -1 &   -26 &   -22&    22 &   -62 &   -46 &     7 &     0 &   -26 &    -2 &   -11\\
 2.00 & 4740 &     9 &   -17 &     6 &    -2 &   -18 &   -18&    19 &   -62 &   -52 &     1 &    -4 &   -36 &    21 &   -32\\
 1.65 & 9211 &    26 &  -118 &     0 &   -35 &   136 &  -218&   -17 &   -60 &   -80 &   -49 &    -6 &     2 &    51 &   -35\\
 1.25 & 8643 &    32 &  -104 &     5 &   -28 &   160 &  -179&    -8 &   -52 &   -65 &   -40 &    -8 &     5 &   -68 &    23\\
 0.50 & 6691 &    23 &   -42 &     7 &    -7 &    89 &   -53&     5 &   -28 &   -34 &    -9 &     0 &     1 &   -36 &    18\\
-0.25 & 5464 &    30 &   -25 &    16 &     0 &    49 &   -38&    13 &   -56 &   -41 &    -3 &    -2 &    -1 &   -49 &    10\\
-1.00 & 4239 &    31 &     0 &    15 &     3 &    26 &   -35&    14 &   -27 &    -3 &    -1 &   -17 &    -2 &    -9 &     4\\
-1.75 & 3643 &   107 &    75 &    13 &     5 &   -10 &   -69&    28 &     8 &    20 &     7 &    -9 &   -11 &    15 &   -25
\enddata
\tablecomments{$\Delta \Teff$(mix) = $\Teff$(mix) -- $\Teff$(scaled-solar); superscript $a$ refers to constant Z=0.0188 while subscript $b$ refers to constant $\feh$=0}
\end{deluxetable}

%TABLE 2
\begin{deluxetable}{cccccccccccccccc}
\tablecolumns{16}
\tablewidth{0pc}
\tablecaption{Differences in the H-R Diagram at 1 Gyr\label{tab2}}
\tablehead{
\multicolumn{2}{c}{Scaled-Solar}&\multicolumn{11}{c}{$\Delta \Teff$ (K) at Constant Log(L/$\Ls$)}\\
\colhead{Log(L/$\Ls$)}&\colhead{$\Teff$ (K)}&\colhead{C$^a$}&\colhead{C$^b$}&\colhead{N$^a$}&\colhead{N$^b$}&\colhead{O$^a$}&\colhead{O$^b$}&\colhead{Ne}&\colhead{Mg}&\colhead{Si}&\colhead{S}&\colhead{Ca}&\colhead{Ti}&\colhead{Fe}&\colhead{$\alpha$}}
\startdata
%L/Lo   Teff      CA       Cb     Na      Nb      Oa      Ob     Ne      Mg      Si       S      Ca      Ti      Fe   alpha
 2.30 & 4328 &    15 &   -14 &    11 &     4 &   -32 &   -26 &    23 &   -62 &   -46 &     5 &     0 &   -24 &     4 &   -18\\
 2.00 & 4549 &    15 &   -13 &    12 &     5 &   -34 &   -23 &    23 &   -60 &   -45 &     6 &     0 &   -27 &     4 &   -22\\
 1.25 & 7705 &    22 &   -96 &     5 &   -43 &    80 &  -209 &   -20 &   -46 &   -62 &   -29 &     2 &     5 &    40 &   -56\\
 0.50 & 6641 &    21 &   -48 &     9 &    -7 &    93 &   -60 &    -3 &   -33 &   -45 &   -19 &    -1 &     0 &   -43 &    20\\
-0.25 & 5455 &    31 &   -24 &    17 &     1 &    52 &   -37 &    14 &   -56 &   -40 &    -2 &    -1 &     0 &   -49 &    10\\
-1.00 & 4233 &    30 &     9 &    12 &     3 &    25 &   -39 &    13 &   -27 &    -4 &    -2 &   -17 &    -2 &    -9 &    11\\
-1.75 & 3641 &   107 &    73 &    13 &     5 &    -4 &   -70 &    27 &     7 &    20 &     7 &   -11 &   -11 &    15 &   -25\\
-2.50 & 3513 &    87 &    62 &    15 &     5 &   -28 &  -169 &    51 &    30 &    29 &    15 &    12 &    12 &    24 &   -68
\enddata
\end{deluxetable}

%TABLE 3
\begin{deluxetable}{cccccccccccccccc}
\tablecolumns{16}
\tablewidth{0pc}
\tablecaption{Differences in the H-R Diagram at 2 Gyr\label{tab3}}
\tablehead{
\multicolumn{2}{c}{Scaled-Solar}&\multicolumn{11}{c}{$\Delta \Teff$ (K) at Constant Log(L/$\Ls$)}\\
\colhead{Log(L/$\Ls$)}&\colhead{$\Teff$ (K)}&\colhead{C$^a$}&\colhead{C$^b$}&\colhead{N$^a$}&\colhead{N$^b$}&\colhead{O$^a$}&\colhead{O$^b$}&\colhead{Ne}&\colhead{Mg}&\colhead{Si}&\colhead{S}&\colhead{Ca}&\colhead{Ti}&\colhead{Fe}&\colhead{$\alpha$}}
\startdata
%L/Lo   Teff      Ca       Cb     Na      Nb      Oa      Ob     Ne      Mg      Si       S      Ca      Ti      Fe   alpha
 3.10 & 3645 &   43 &     1 &    14 &     3 &   -14 &   -35  &    38 &   -72 &   -18 &    13 &   -14 &    -6 &     2 &    -5\\
 2.75 & 3907 &   24 &   -11 &    12 &     2 &   -11 &   -20  &    34 &   -76 &   -22 &    12 &   -10 &    -4 &    -5 &     1\\
 2.00 & 4433 &   13 &   -27 &    10 &    -1 &   -13 &   -16  &    33 &   -86 &   -40 &    12 &    -2 &    -3 &   -15 &     0\\
 1.25 & 4904 &   17 &   -25 &    12 &     0 &   -18 &   -19  &    32 &   -92 &   -53 &    11 &    -3 &    -5 &     0 &   -13\\
 0.80 & 6703 &    4 &   -63 &     9 &   -21 &    61 &  -105  &   -14 &   -54 &   -38 &    -5 &   -16 &    -4 &     0 &    -6\\
 0.50 & 6501 &   -1 &   -32 &     5 &    -8 &    63 &   -43  &    -2 &   -32 &   -35 &    -7 &     0 &    -1 &   -37 &    11\\
-0.25 & 5441 &   30 &   -24 &    16 &     0 &    58 &   -38  &    14 &   -58 &   -41 &    -2 &    -2 &    -1 &   -50 &    10\\
-1.00 & 4227 &   31 &     1 &    13 &     3 &    27 &   -40  &    14 &   -28 &    -3 &    -1 &   -18 &    -2 &    -8 &     0\\
-1.75 & 3645 &   97 &    62 &    12 &    -1 &   -12 &   -75  &    19 &     0 &    11 &     0 &   -11 &    -9 &     6 &   -25\\
-2.50 & 3513 &   86 &    62 &    15 &     5 &   -27 &  -168  &    51 &    30 &    29 &    15 &    12 &    12 &    24 &   -67
\enddata
\end{deluxetable}

%TABLE 4
\begin{deluxetable}{cccccccccccccccc}
\tablecolumns{16}
\tablewidth{0pc}
\tablecaption{Differences in the H-R Diagram at 4 Gyr\label{tab4}}
\tablehead{
\multicolumn{2}{c}{Scaled-Solar}&\multicolumn{11}{c}{$\Delta \Teff$ (K) at Constant Log(L/$\Ls$)}\\
\colhead{Log(L/$\Ls$)}&\colhead{$\Teff$ (K)}&\colhead{C$^a$}&\colhead{C$^b$}&\colhead{N$^a$}&\colhead{N$^b$}&\colhead{O$^a$}&\colhead{O$^b$}&\colhead{Ne}&\colhead{Mg}&\colhead{Si}&\colhead{S}&\colhead{Ca}&\colhead{Ti}&\colhead{Fe}&\colhead{$\alpha$}}
\startdata
%L/Lo   Teff      Ca       Cb     Na      Nb      Oa      Ob     Ne      Mg      Si       S      Ca      Ti      Fe   alpha
 3.30 & 3347 &    60 &    18 &    17 &     2 &   -12 &   -49  &    43 &   -75 &   -13 &    18 &   -15 &    -7 &     4 &    -7\\
 2.75 & 3799 &    36 &     3 &    15 &     3 &    -5 &   -24  &    37 &   -73 &   -16 &    15 &   -12 &    -4 &    -4 &     6\\
 2.00 & 4311 &    13 &   -21 &    12 &    -2 &    -2 &   -14  &    33 &   -83 &   -32 &    14 &    -1 &    -1 &   -19 &    11\\
 1.25 & 4745 &    18 &   -22 &    14 &    -2 &    -3 &   -20  &    35 &   -90 &   -42 &    14 &    -2 &    -1 &   -25 &    11\\
 0.50 & 6139 &    13 &   -50 &     7 &    -8 &    49 &   -66  &     3 &   -48 &   -46 &    -4 &    -2 &    -2 &   -61 &    11\\
-0.25 & 5413 &    31 &   -23 &    16 &     0 &    68 &   -37  &    15 &   -60 &   -41 &     0 &    -2 &    -1 &   -50 &    10\\
-1.00 & 4221 &    31 &     2 &    12 &     3 &    28 &   -41  &    14 &   -29 &    -3 &    -2 &   -19 &    -3 &    -9 &    -4\\
-1.75 & 3635 &   100 &    66 &    14 &     3 &     1 &   -72  &    23 &     5 &    16 &     4 &    -5 &    -5 &    12 &   -21\\
-2.50 & 3513 &    86 &    62 &    15 &     5 &   -27 &  -167  &    51 &    29 &    29 &    15 &    12 &    12 &    24 &   -67
\enddata
\end{deluxetable}

%TABLE 5
\begin{deluxetable}{cccccccccccccccc}
\tablecolumns{16}
\tablewidth{0pc}
\tablecaption{Differences in the H-R Diagram at 8 Gyr\label{tab5}}
\tablehead{
\multicolumn{2}{c}{Scaled-Solar}&\multicolumn{11}{c}{$\Delta \Teff$ (K) at Constant Log(L/$\Ls$)}\\
\colhead{Log(L/$\Ls$)}&\colhead{$\Teff$ (K)}&\colhead{C$^a$}&\colhead{C$^b$}&\colhead{N$^a$}&\colhead{N$^b$}&\colhead{O$^a$}&\colhead{O$^b$}&\colhead{Ne}&\colhead{Mg}&\colhead{Si}&\colhead{S}&\colhead{Ca}&\colhead{Ti}&\colhead{Fe}&\colhead{$\alpha$}}
\startdata
%L/Lo   Teff      Ca       Cb     Na      Nb      Oa      Ob     Ne      Mg      Si       S      Ca      Ti      Fe   alpha
 3.30 & 3244 &   61 &    10 &    15 &     2 &   -24 &   -70  &    43 &   -74 &   -12 &    17 &   -15 &    -9 &    -2 &   -10\\
 2.75 & 3718 &   42 &    -2 &    15 &     3 &   -16 &   -44  &    38 &   -69 &   -13 &    15 &   -12 &    -5 &    -6 &     5\\
 2.00 & 4233 &   11 &   -30 &    10 &    -1 &   -16 &   -30  &    32 &   -80 &   -28 &    13 &    -2 &    -2 &   -22 &    11\\
 1.25 & 4662 &   15 &   -29 &    11 &    -1 &   -19 &   -34  &    34 &   -85 &   -36 &    14 &    -1 &    -3 &   -30 &    10\\
 0.50 & 4969 &   23 &   -43 &    15 &    -3 &    35 &   -43  &    37 &  -114 &   -51 &    22 &     1 &    -2 &   -33 &    13\\
 0.30 & 5871 &   17 &   -36 &     7 &    -8 &    77 &   -48  &     7 &   -65 &   -43 &     9 &    -1 &    -3 &   -36 &    10\\
-0.25 & 5359 &   30 &   -22 &    15 &     0 &    83 &   -35  &    14 &   -63 &   -41 &     1 &    -2 &    -1 &   -51 &    10\\
-1.00 & 4211 &   33 &     5 &    12 &     4 &    35 &   -41  &    15 &   -29 &    -2 &    -1 &   -20 &    -3 &    -8 &    -4\\
-1.75 & 3629 &  102 &    68 &    12 &     5 &     1 &   -70  &    26 &     7 &    18 &     6 &    -9 &    -8 &    14 &   -26\\
-2.50 & 3513 &   86 &    62 &    14 &     4 &   -20 &  -167  &    51 &    29 &    28 &    15 &    11 &    11 &    24 &   -67
\enddata
\end{deluxetable}

%TABLE 6
\begin{deluxetable}{cccccccccccccccc}
\tablecolumns{16}
\tablewidth{0pc}
\tablecaption{Differences in the H-R Diagram at 12 Gyr\label{tab6}}
\tablehead{
\multicolumn{2}{c}{Scaled-Solar}&\multicolumn{11}{c}{$\Delta \Teff$ (K) at Constant Log(L/$\Ls$)}\\
\colhead{Log(L/$\Ls$)}&\colhead{$\Teff$ (K)}&\colhead{C$^a$}&\colhead{C$^b$}&\colhead{N$^a$}&\colhead{N$^b$}&\colhead{O$^a$}&\colhead{O$^b$}&\colhead{Ne}&\colhead{Mg}&\colhead{Si}&\colhead{S}&\colhead{Ca}&\colhead{Ti}&\colhead{Fe}&\colhead{$\alpha$}}
\startdata
%L/Lo   Teff      Ca       Cb     Na      Nb      Oa      Ob     Ne      Mg      Si       S      Ca      Ti      Fe   alpha
 3.30 & 3174 &    65 &    15 &    15 &     1 &   -19 &   -66 &    44 &   -75 &   -12 &    17 &   -16 &    -9 &    -2 &    -7\\
 2.75 & 3663 &    46 &     2 &    14 &     3 &   -13 &   -45 &    39 &   -70 &   -13 &    15 &   -13 &    -6 &    -6 &     4\\
 2.00 & 4178 &    12 &   -25 &    10 &    -1 &   -10 &   -25 &    33 &   -79 &   -26 &    14 &    -2 &    -2 &   -21 &    12\\
 1.25 & 4599 &    15 &   -26 &    11 &    -2 &   -12 &   -28 &    35 &   -86 &   -35 &    14 &    -1 &    -3 &   -30 &    12\\
 0.50 & 4844 &    19 &   -27 &    13 &    -1 &    23 &   -22 &    37 &   -96 &   -39 &    16 &    -2 &    -2 &   -37 &    15\\
 0.15 & 5649 &    20 &   -33 &     9 &    -6 &    92 &   -37 &    12 &   -74 &   -46 &     9 &    -1 &    -2 &   -41 &    10\\
-0.25 & 5305 &    29 &   -22 &    14 &    -1 &    92 &   -31 &    13 &   -66 &   -40 &     2 &    -3 &    -1 &   -50 &     9\\
-1.00 & 4204 &    34 &     5 &    12 &     4 &    38 &   -42 &    15 &   -31 &    -3 &    -1 &   -22 &    -4 &    -8 &    -6\\
-1.75 & 3628 &   102 &    68 &    12 &     5 &    -1 &   -69 &    26 &     7 &    19 &     7 &    -9 &   -10 &    14 &   -26\\
-2.50 & 3513 &    85 &    61 &    14 &     4 &   -21 &  -165 &    50 &    28 &    28 &    15 &    11 &    11 &    23 &   -67

\enddata
\end{deluxetable}

\clearpage
\subsection{Surface Boundary Conditions}
Figure \ref{iso6} compares the difference between scaled-solar and $\alpha$-enhanced using the Eddington T-$\tau$ boundary condition on the left and PHOENIX model atmosphere boundary condition on the right. The composition in the model atmosphere boundary condition has been set to that of the stellar evolution models.  The mixing length used to calibrate the Eddington T-$\tau$ models has been used in the PHOENIX models as well and thus the PHOENIX isochrones are meant to be compared only to each other and not to the Eddington isochrones. Independent of the mixing length, the use of different boundary conditions will result in a different absolute temperature scale in the models.  However, it is possible to ask whether or not the temperature difference at constant luminosity between one composition and another is dependent upon the choice of boundary condition. In a differential sense, as Figure \ref{iso6} demonstrates, there is good agreement between the Eddington and PHOENIX boundary conditions in the predicted temperature offsets. This suggests that the differential results presented in this paper are robust at least to changes in the surface boundary condition.

\subsection{Main sequence lifetimes}
\clearpage
%Figure 13
\begin{figure}
\epsscale{1.0}
\plotone{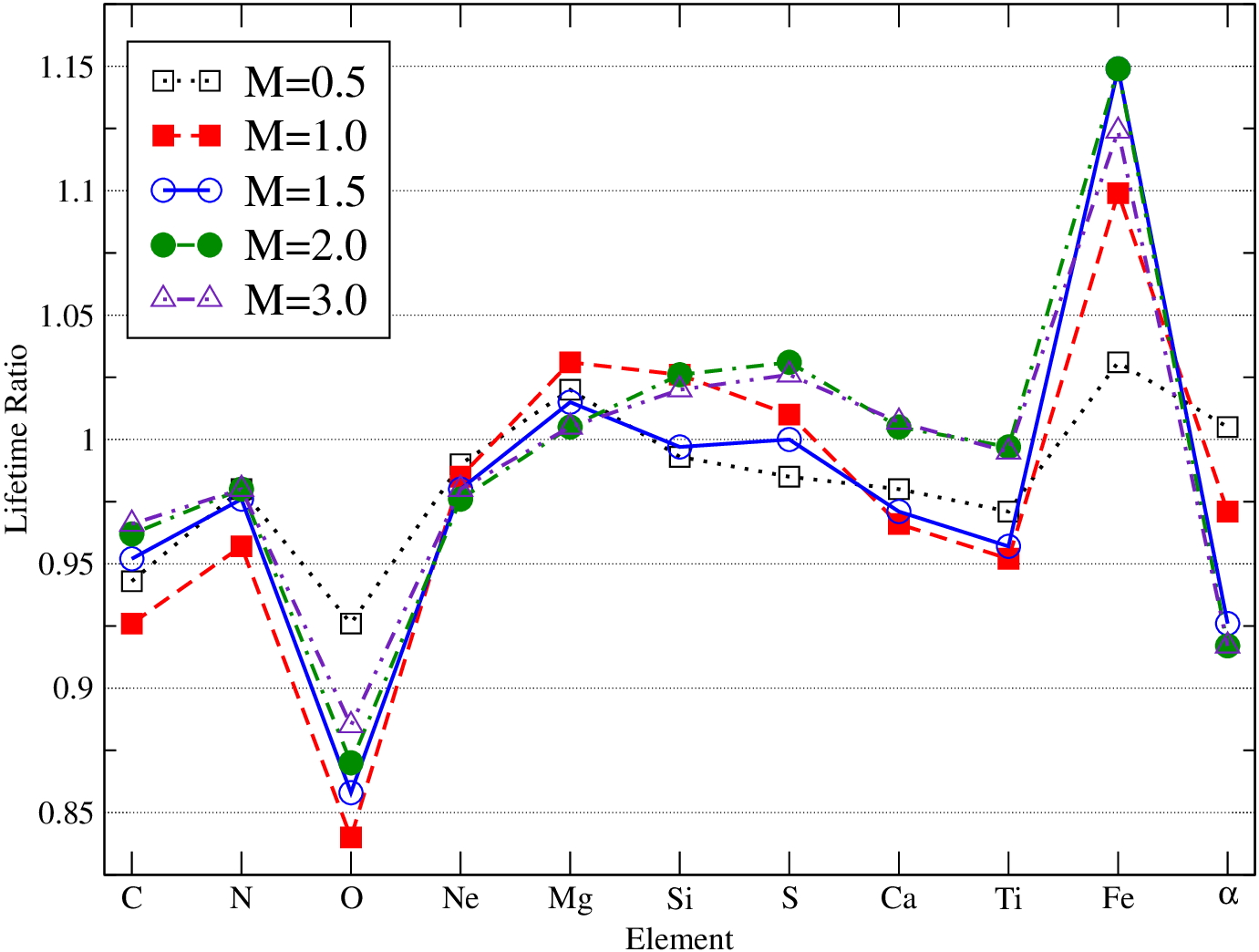}
\caption{The influence of individual elements and $\alpha$-enhancement on main sequence lifetimes relative to scaled-solar for five different masses. The numbers plotted are the ratio of main sequence lifetime of an enhanced element mixture to scaled-solar for a given mass: lifetime(mix)/lifetime(scaled-solar). O and Fe have the most dramatic effect on main sequence lifetimes. For clarity, the O-enhanced case at $\feh$=0 is not shown. There does not appear to be any significant trend with stellar mass.  Masses listed in the legend are in solar units.\label{lifetimes}}
\end{figure}

Main sequence lifetimes\footnote{Defined as the time elapsed between the zero age main sequence (when the H-burning luminosity reaches 99\% of the total luminosity) and core H exhaustion.} for the individually-enhanced elements and $\alpha$-enhancement have been compared to scaled-solar for five stellar masses: 0.5, 1, 1.5, 2, and 3 $\Ms$. Ratios in the figure are lifetime(mix)/lifetime(scaled-solar).  Most elements change stellar lifetimes by 5\% or less on average, regardless of mass. O- and Fe-enhanced tracks have main sequence lifetimes that differ from scaled-solar by roughly -10\% and +10\% respectively for masses of at least 1 $\Ms$.  Figure \ref{lifetimes} demonstrates the lifetime ratios with respect to scaled-solar for the five masses listed above. There are no obvious trends with mass nor with low temperature opacities.  However, there does appear to be a relationship between lifetime ratios and high temperature opacities.  Compare Figure \ref{lifetimes} to the lower panels of Figures \ref{opac3} and \ref{opac4}. O and Fe show the largest deviations in opacity between log T=5 and 8 (see $\S$4.1.2) compared to scaled-solar. The most obvious explanation is that depleting or enhancing Fe (and the heavier $\alpha$-elements) abundance reduces or increases mean opacity at high temperatures, respectively.

In addition to changes in the opacities, changes to C, N, and O will alter the amount of energy generated via the CNO cycle.  To account for this possibility, test cases were computed at constant Z for several masses with enhanced C, N, or O but using the scaled-solar opacities. The effect will obviously be greater in more massive stars where the CNO cycle plays a more important role on the main sequence. At 1 $\Ms$, the scaled-solar model has a slightly longer main sequence lifetime than the C-, N-, or O-enhanced models but the difference is at most 0.2\%.  At 3 $\Ms$ the scaled-solar model has a shorter lifetime but the effect is small: C increases lifetime by 0.5\%, N increases lifetime by less than 0.1\%, while O leaves the lifetime essentially unchanged. At 5 $\Ms$ the scaled-solar model again has a shorter lifetime but the largest difference is only 0.5\%. By comparison, including the appropriate opacities, the C-, N-, and O-enhanced models are $\sim$5\% or more younger than their scaled-solar counterparts for all masses (see Figure \ref{lifetimes}).

\section{Discussion}
As a general rule, increasing opacity should result in a cooler, longer-lived star but the previous section suggests the issue is more complex. Low temperature opacities have more influence on the effective temperature scale while high temperature opacities have more influence on lifetimes. Unfortunately, it is not possible to derive a more concrete statement about the opacities that explain everything presented in the previous section.

Enhancing O at constant Z causes the mean opacity to decrease because O is replacing substantial amounts of the heavier elements that tend to increase opacity. The same is true for C, N, and Ne.  These elements also generate isochrones with hotter effective temperatures than scaled-solar.  The trend for these elements is to decrease main sequence lifetimes relative to scaled-solar with O causing the greatest reduction of any element and Ne having the least effect of all the elements regardless of mass.  The coolest stellar models included in this study barely reach temperatures where absorption from molecules containing C and O make major contributions.

Relaxing the constant Z constraint by enhancing C, N, and O at $\feh$=0 results in stellar models and isochrones that are generally cooler than their scaled-solar counterparts.  However, C-enhanced isochrones are still hotter than scaled-solar on the lower main sequence (below 4,000 K). N remains a minor influence on the effective temperature scale. Raising $\feh$ to zero with enhanced O requires a 50\% increase in total Z (0.0188 to 0.0276) and produces tracks and isochrones that are consistently cooler and longer-lived than at $\feh$=0 for scaled-solar Z.

Enhancing Mg and Si shift isochrones to cooler temperatures but lifetimes increase by about 3\% at most.  Enhancing both elements lead to increased low and high temperature opacities.  With Mg more influential at low temperature and Si more so at high temperature. It is hard to differentiate between the importance of low and high temperature opacities on stellar models for these two elements. The findings presented here underline the importance of accurately determining Mg and Si abundances, in addition to Fe, in stellar populations.

Enhancing S at constant Z causes the low temperature opacities to decrease and the high temperature opacities to increase relative to scaled-solar.  The effective temperature scale is essentially unchanged from scaled-solar in most cases.  The lifetimes are slightly shorter for M $<$ 1 $\Ms$ and slightly higher otherwise.

Ca and Ti are the least abundant elements considered.  Each reduces lifetimes by less than 5\% regardless of stellar mass.  They become more important as the temperature decreases, becoming as influential as Si on the red giant branch for the oldest ages.  Oddly, though enhancing either Ca or Ti shifts the main sequence to cooler temperatures than scaled-solar, these two elements also slightly decrease main sequence lifetimes.

Enhancing Fe clearly causes the effective temperatures to fall and the lifetimes to increase significantly.  Fe increases both low and high temperature opacities and so it is no surprise that lifetimes increase and effective temperatures decrease relative to scaled-solar. One unusual result of Fe-enhancement is the increased luminosity of the main sequence turn off and sub-giant branch at younger ages (Figure \ref{iso5}). This effect is due to the lengthening of lifetimes by Fe. Fe-enhancement significantly increases opacity at core temperatures and thereby inhibits nuclear reactions by cooling the core. This effect is more pronounced in more massive stars but is present in all cases.  Stellar mass at the main sequence turn off and on the subgiant branch is higher than for scaled-solar at a given age and therefore more luminous. Because Fe also shifts the tracks to cooler temperatures the Fe-enhanced isochrones are more brighter but not much hotter (often cooler) than scaled-solar isochrones at the same age.

Enhancing the $\alpha$-elements in the aggregate is qualitatively similar to enhancing O but to a lesser extent. The lessening is likely due to the fact that while enhancing O substantially depletes other elements, some of those are put back (especially Mg and Si). On the other hand, enhancing the $\alpha$-elements severely depletes Fe, thus causing a reversal of the effect observed in Fe-enhancement.  Figure \ref{iso6} shows that at 1 Gyr the $\alpha$-enhanced isochrone has a fainter main sequence turn off and sub-giant branch. With Fe-enhancement, the turnoff and sub-giant branch become brighter and we see that a depletion of Fe ($\alpha$-enhancement has the lowest $\feh$ of all mixtures, see Figure \ref{feh}) causes the these features to become fainter.  This effect is not obvious in any of the other isochrone comparisons and deserves further attention.

\section{Conclusions}
Isochrones constructed with twelve different heavy element mixtures, all held at constant overall Z, were analyzed to determine how an individual element contributes to the evolutionary properties of stellar models.  The twelve sets include scaled-solar, alpha-enhanced, and ten distinct sets where C, N, O, Ne, Mg, Si, S, Ca, Ti, and Fe were enhanced individually.  By comparing stellar evolution tracks and isochrones, a picture of how each element influences the evolution has begun to emerge.

An element whose main contribution to opacity occurs at low temperatures should alter the effective temperature scale more than it alters stellar lifetimes and the opposite should occur for an element that contributes more opacity at high temperature.  In practice, the elements contribute differently over a wide range of temperatures and must be evaluated individually or in small groups.  Enhancing the lighter elements (C, N, O) at constant Z produces hotter, shorter-lived stars. The heavier elements do not group together well.  Enhancing Fe produce cooler, longer-lived stars. Enhancing Mg, Si, S, Ca, and Ti generally produces cooler stars but the lifetimes can either increase or decrease by a few percent.  For temperatures of Log T=3.5 and above, the lighter elements contribute more by displacing the heavier elements.

This study is by no means complete in the sense that it fully addresses all possible issues related to abundance variations and stellar evolution.  It is, however, a useful exercise because it demonstrates that in some cases substantial changes do occur and effort ought to be made to understand and account for them.

In a companion paper (Lee et al. 2007, in preparation) the analysis will be extended to spectroscopic properties and integrated light models. Isochrones covering a wider range of X and Z with individual element enhancements are currently in production.

\acknowledgments
Support for this work was provided by the NSF through grant AST-0307487, the New Standard Stellar Population Model (NSSPM) project.

AD thanks Achim Weiss for reading the manuscript and helping to verify some of the results.

JF acknowledges support from NSF grant AST-0239590 and Grant No. EIA-0216178, Grant No. EPS-0236913 with matching support from the State of Kansas and the Wichita State University High Performance Computing Center.

The work of DJ and EB was supported in part by by NASA grants NAG5-3505 and NAG5-12127, and NSF grant AST-0307323. This research used resources of the National Energy Research Scientific Computing Center (NERSC), which is supported by the Office of Science of the U.S.  Department of Energy under Contract No. DE-AC03-76SF00098; and the H\"ochstleistungs Rechenzentrum Nord (HLRN).  We thank both these institutions for a generous allocation of computer time.

DJ thanks the Ministry of Science and Environment Protection of Serbia for support received from Project No. 146001.


\begin{thebibliography}{}
\bibitem[Alexander \& Ferguson(1994)]{alex}Alexander, D. R. \& Ferguson, J. W. 1994, \apj, 437, 879
\bibitem[Bahcall et al.(2005)]{bahc}Bahcall, J. N., Basu, S., Pinsonneault, M., \& Serenelli, A. M. 2005, \apj, 618,1049
\bibitem[Bjork \& Chaboyer(2006)]{bjor}Bjork, S. R. \& Chaboyer, B. 2006, \apj, 641, 1102
\bibitem[Chaboyer et al.(2001)]{chab}Chaboyer, B., Fenton, W. H., Nelan, J. E., Patnaude, D. J., \& Simon, F. E. 2001, \apj, 562, 521
\bibitem[Chieffi et al.(1991)]{chie} Chieffi, A., Straniero, O., \& Salaris, M. 1991, ASPC, 13, 219
\bibitem[Demarque et al.(2004)]{dema}Demarque, P., Woo, J.-H., Kim, Y.-C., \& Yi, S. K. 2004, \apjs, 155, 667
\bibitem[Ferguson et al.(2005)]{ferg}Ferguson, J.~W., Alexander, D.~R., Allard, F., Barman, T., Bodnarik, J.~G., Hauschildt, P.~H., Heffner-Wong, A., \& Tamanai, A. 2005 \apj, 623, 585
\bibitem[Grevesse \& Sauval(1998)]{gs98}Grevesse, N. \& Sauval, A. J. 1998, \ssr, 85, 161
\bibitem[Hauschildt et al.(1999a)]{phxa}Hauschildt, P. H., Allard, F., \& Baron, E. 1999a, \apj, 512, 377 
\bibitem[Hauschildt et al.(1999b)]{phxb}Hauschildt, P. H., Allard, F., Ferguson, J., Baron, E., \& Alexander, D. 1999b, \apj, 525, 871
\bibitem[Iglesias \& Rogers(1996)]{opal}Iglesias, C.A. \& Rogers, F.J. 1996, \apj, 464, 943
\bibitem[Irwin(2004)]{irwin}Irwin, A. 2004, http://freeeos.sourceforge.net/eff\_fit.pdf
\bibitem[Kim et al.(2002)]{kim}Kim, Y.-C., Demarque, P., Yi, S. K., Alexander, D. R. 2002, \apjs, 143, 499
\bibitem[Korn et al.(2005)]{korn}Korn, A. J., Maraston, C., Thomas, D. 2005 \aap, 438, 685
\bibitem[Pietrinferni et al.(2006)]{piet}Pietrinferni, A., Cassisi, S., Salaris, M., \& Castelli, F.  2006, \apj, 642, 797
\bibitem[Salasnich et al.(2000)]{salasnich}Salasnich, B., Girardi, L., Weiss, A., \& Chiosi, C. 2000, \aap, 361, 1023
\bibitem[Seaton et al.(1994)]{seat}Seaton, M. J., Yan, Y., Mihalas, D., \& Pradhan, A. K. 1994, \mnras, 266, 805
\bibitem[Serven et al.(2005)]{serv}Serven, J., Worthey, G., \& Briley, M. M. 2005, \apj, 627, 754
\bibitem[Sestito et al.(2006)]{degl}Sestito, P., Degl'Innocenti, S., Prada Moroni, P. G., Randich, S. 2006, \aap, 454, 311
\bibitem[Tripicco \& Bell(1995)]{trip}Tripicco, M. J. \& Bell, R. A. 1995, \aj, 110, 3035
\bibitem[Vandenberg(1992)]{vand}Vandenberg, D. A. 1992, \apj, 391, 685
\bibitem[Vandenberg \& Bell(2001)]{vandb}Vandenberg, D. A. \& Bell, R. A. 2001, \nar, 45, 577
\bibitem[Vandenberg et al.(2000)]{vand2}VandenBerg, D. A., Swenson, F. J., Rogers, F. J., Iglesias, C. A., Alexander, D. R. 2000, \apj, 532, 430
\bibitem[Weiss et al.(2006)]{weiss} Weiss, A., Salaris, M., Ferguson, J. W., and Alexander, D. R. 2006, \aap, submitted (astro-ph/0605666)
\end{thebibliography}
\end{document}